\documentclass[aps,pra,nobibnotes,showpacs,amsfonts,amsmath,amssymb,onecolumn]{revtex4}

\usepackage{graphicx}

\begin{document}

\title{Entanglement in the 
dispersive interaction of trapped ions with a quantized field}
    \author{F. L. Semi\~ao}
    \author{K. Furuya}
        \affiliation{Instituto de F\'\i sica `Gleb Wataghin',
        Universidade Estadual de Campinas,  13083-970 Campinas, SP,
        Brazil}
\date{\today}

\begin{abstract}
The mode-mode entanglement between trapped ions and cavity fields is investigated in the dispersive regime. We show how a simple initial preparation of Gaussian coherent states and a postselection may be used to generate motional non-local mesoscopic states (NLMS) involving ions in different traps. We also present a study of the entanglement induced by dynamical Stark-shifts considering a cluster of N-trapped ions. In this case, all entanglement is due to the dependence of the Stark-shifts on the ions' state of motion manifested as a cross-Kerr interaction between each ion and the field.
\end{abstract}
\maketitle
\section{introduction}
When thinking of practical implementation of quantum computers and protocols, two types of information carriers have been widely considered. The first type is composed of two-level systems (qubits) such as quantum dots \cite{qdots} or superconducting devices \cite{super}, and the second type is composed of continuous variable CV systems like the continuous quadrature amplitudes of electromagnetic fields \cite{cv}. These two types of quantum information carriers are usually thought as unit blocks (flying and stationary qubits) for distributed quantum computation \cite{dist}. For long distance communication, photons are more appropriate to carry information between stationary qubits and are labeled flying qubits. It is then clear that in order to fully explore the distributed quantum computation content, it is important to investigate the interaction between CV systems and qubits. In this context, trapped ions interacting with quantized electromagnetic fields \cite{zl,peter,ticqed,ticqedexp} constitutes a promising physical system for it comprises the interaction of a finite system (internal electronic levels) with two bosonic modes namely the cavity field and the vibrational center-of-mass motion. 
Distributed quantum computation relies heavily on entanglement which is one of the key ingredients for the achievement of quantum information tasks. For instance, when supplied with a pair of maximally entangled states, two parties may achieve perfect teleportation \cite{bennett1} and dense coding \cite{bennett2} using local operations and classical communications (LOCC). Such applications require entanglement as a resource and clearly show the importance of studying it in different physical systems for actual implementation.

In this work, we study the entanglement properties of a cluster of two-level trapped ions interacting with a quantized electromagnetic field. This multipartite system has been first considered by Bu\v{z}ek \textit{et al} in \cite{peter} where they explore a special resonance condition allowing the system to be effectively described as a collective Tavis-Cummings model \cite{tv}. Here, we consider the opposite situation where the cavity and the internal electronic levels are kept far off resonance with each other so that the field interacts only dispersively with the ions. This paper is organized as follows. In Section \ref{m} we describe the dispersive interaction of N-trapped ions with a quantized electromagnetic field. In Section \ref{gen} we show that depending on the initial state of cluster and field, the cavity induced atomic dipole-dipole couplings may lead to the generation of non-local mesoscopic states (NLMS) \cite{entcoh,entprop,haroche}. These states are known to violate Bell inequalities for suitable choices of sampling points in phase space \cite{desbell}, and in our generation protocol, they are naturally protected from destructive effects of cavity losses. The NLMS generation is achieved after a simple measurement of the internal state of the ions (postselection). Section \ref{sta} is dedicated to the study of mode-mode entanglement induced by motion-dependent Stark-shifts \cite{kerr} arising from the dispersive interaction with the field. In this situation, we analyze the dynamics of entanglement between the elements in the cluster of ions or between field and cluster, and compare these bipartite quantities to the multipartite global entanglement measure \cite{global} whose evaluation involves all ions and the cavity field. The ideas and effects discussed here in terms of trapped ions could, in principle, be extended to neutral atoms trapped in optical lattices \cite{lattices}. The standing wave laser beams possess the periodic structure able to trap individual atoms the way it is needed to the applications shown in this paper. Of course, an additional experimental challenge would be to put such atoms in interaction with a quantized field, typically a cavity mode. Section \ref{concl} is devoted to some conclusions and final comments. 
\section{Model}\label{m}
We consider a cluster of N-trapped ions placed in a resonator which sustains a particular mode of electromagnetic field. The whole system is treated quantum mechanically and its Hamiltonian is given by($\hbar=1$) \cite{ticqed}
\begin{eqnarray}\label{H}
\hat{H}=\sum_{i=1}^N\nu_i\hat{a}_i^\dag\hat{a}_i+\omega_c\hat{b}^\dag\hat{b}+\sum_{i=1}^N\omega_i\frac{\hat{\sigma}_i^z}{2}+\sum_{i=1}^N g_i\cos[\eta_i(\hat{a}_i^\dag+\hat{a}_i)](\hat{\sigma}_i^+\hat{b}+\hat{\sigma}_i^-\hat{b}^\dag),
\end{eqnarray}
where $\hat{a}_i(\hat{a}_i^\dag)$ denotes the annihilation (creation) operator for the vibrational mode of center-of-mass motion of ion $i$ (frequency $\nu_i$), $\hat{b}_i(\hat{b}_i^\dag)$ denotes the annihilation (creation) operator for cavity field mode (frequency $\omega_c$), $\hat{\sigma}_i$'s are the usual
Pauli atomic operators for the two-level trapped ion $i$, $g_i$ is the field-ion $i$ coupling constant, and $\eta_i=2\pi
\Delta x_i/\lambda$ is the Lamb-Dicke parameter, $\Delta x_i$ being the rms fluctuation of the position of ion $i$ in its lowest trap eigenstate, and $\lambda$ the wavelength of the cavity field. Also, we are considering that the traps are sufficiently far apart so that one can neglect the Coulomb repulsion between the ions.

In what follows, we will be interested in the cross-Kerr coupling of the bosonic field and motion operators. This is derived in detail in \cite{kerr} for the case $N=1$ where the dispersive interaction of ion-field has been considered. We are now going to derive the correspondent effective Hamiltonian accounting for the dispersive interaction of the cavity field and the cluster. This may be achieved by a careful analysis of the quantum equations of motion and application of the high detuning condition $\Delta_i\equiv\omega_i-\omega_c\gg g_i$. We first note that the general Hamiltonian (\ref{H}) is highly nonlinear
because the function $\cos\eta(\hat{a}_i^{\dagger}+\hat{a}_i)$ contains powers of combinations of bosonic operators. It is shown in \cite{kerr,tp} that in the regime $\Delta_i\ll \nu_i$, it is possible to keep just the terms proportional to powers of the number operator $\hat{a}_i^{\dagger}\hat{a}_i$ so that the Hamiltonian (\ref{H}) may be approximated by
\begin{eqnarray}\label{Hap}
\hat{H}=\sum_{i=1}^N\nu_i\hat{a}_i^\dag\hat{a}_i+\omega_c\hat{b}^\dag\hat{b}+\sum_{i=1}^N\omega_i\frac{\hat{\sigma}_i^z}{2}+\sum_{i=1}^N g_if(\hat{a}_i^\dag\hat{a}_i)(\hat{\sigma}_i^+\hat{b}+\hat{\sigma}_i^-\hat{b}^\dag),
\end{eqnarray}
where
\begin{eqnarray}
f(\hat{a}^\dag\hat{a})=
e^{-\eta^2/2}:J_0(2\eta\sqrt{\hat{a}^\dag\hat{a}}):,
\end{eqnarray}
and $:J_0:$ is the normally ordered zeroth order Bessel function of
the first kind \cite{bessel}. 

The Heisenberg equation of motion for $\hat{\sigma}_j^+\hat{b}$ with $1\leq j\leq N$ when using the system Hamiltonian (\ref{H}) is given by
\begin{eqnarray}\label{eqH1}
i\frac{d}{dt}\hat{\sigma}_j^+\hat{b}= \Delta_j\hat{\sigma}_j^\dag\hat{b}+g_jf(\hat{a}_j^\dag\hat{a}_j)\hat{\sigma}_j^z\hat{b}\hat{b}^\dag+\sum_{i=1}^N g_if(\hat{a}_i^\dag\hat{a}_i)\hat{\sigma}_i^-\hat{\sigma}_j^+.
\end{eqnarray}
It is now convenient to move to an appropriate reference frame by making
$\hat{\sigma}_{j}^+\rightarrow\hat{\sigma}_{j}^+e^{i\omega_jt}$,
$\hat{b}\rightarrow\hat{b}e^{-i\omega_c t}$ and
$\hat{a}_j\rightarrow\hat{a}_je^{-i\nu_j t}$. In this new frame, we may
rewrite (\ref{eqH1}) as
\begin{eqnarray}\label{eqH2}
i\frac{d}{dt}\hat{\sigma}_j^+\hat{b}= g_jf(\hat{a}_j^\dag\hat{a}_j)\hat{\sigma}_j^z\hat{b}\hat{b}^\dag e^{-i\Delta_j t}+ \sum_{i=1}^N g_if(\hat{a}_i^\dag\hat{a}_i)\hat{\sigma}_i^- \hat{\sigma}_j^+e^{-i\Delta_i t}.
\end{eqnarray}
In the limit $\Delta_k\gg g_k$ with $1\leq k\leq N$, we can integrate (\ref{eqH2}) to obtain in the original frame
\begin{eqnarray}\label{sol}
\hat{\sigma}_j^+\hat{b}&=& \frac{g_j}{\Delta_j}f(\hat{a}_j^\dag\hat{a}_j)\hat{\sigma}_j^z\hat{b}\hat{b}^\dag +\sum_{i=1}^N \frac{g_i}{\Delta_i} f(\hat{a}_i^\dag\hat{a}_i)\hat{\sigma}_i^-\hat{\sigma}_j^+. 
\end{eqnarray} 
Now, by substituting (\ref{sol}) in the Hamiltonian (\ref{Hap}), we obtain the effective Hamiltonian \cite{foot1}
\begin{eqnarray}\label{eff}
\hat{H}_{{\rm{eff}}}=\hat{H}_0+\hat{H}_{{\rm{Stark}}}+\hat{H}_{{\rm{dip}}},
\end{eqnarray}
with the free part given by
\begin{equation}
\hat{H}_0=\sum_{i=1}^N\nu_i\hat{a}_i^\dag\hat{a}_i+\omega_c\hat{b}^\dag\hat{b}+\sum_{i=1}^N\omega_i\frac{\hat{\sigma}_i^z}{2},
\end{equation}
the Stark-shifts given by
\begin{equation}\label{st}
\hat{H}_{{\rm{Stark}}}=\sum_{i=1}^N\frac{g_i^2}{\Delta_{i}}f^2(\hat{a}_i^\dag\hat{a}_i)[(1+\hat{b}\hat{b}^\dag)\hat{\sigma}_i^+\hat{\sigma}_i^- -\hat{b}^\dag\hat{b}\hat{\sigma}_i^-\hat{\sigma}_i^+]
\end{equation}
and the energy exchange term takes the form of a atomic dipole-dipole coupling given by
\begin{equation}\label{dip}
\hat{H}_{{\rm{dip}}}=\sum_{{i,j=1} \atop{i\neq j}}^N\frac{g_ig_j}{\Delta_{j}}f(\hat{a}_i^\dag\hat{a}_i)f(\hat{a}_j^\dag\hat{a}_j)(\hat{\sigma}_i^+\hat{\sigma}_j^-+\hat{\sigma}_i^-\hat{\sigma}_j^+).
\end{equation}
The effective Hamiltonian (\ref{eff}) is valid in the regime $\nu_k\gg\Delta_k\gg g_k$, with $1\leq k\leq N$. This effective Hamiltonian describes several interesting physical effects. We note in (\ref{st}) that the dynamical Stark-shifts depend upon the ions' center-of-mass motion via $f(\hat{a}_i^\dag\hat{a}_i)$. Despite the obvious interest this might attract from the point of view of fundamental atomic and optical physics, it turns out to be an unusual way for entangling bosonic modes in a controlled way as we show later in this paper . We may also see that the dipole-dipole Hamiltonian also depends on the intensity of the vibrational motion of the ions via $f(\hat{a}_i^\dag\hat{a}_i)$. In the next section, we explore such Hamiltonian for generating NLMS involving the motional state of ions in different traps. 

It is worth noticing that the Hamiltonian (\ref{eff}) splits into (\ref{st}) or (\ref{dip}) depending on the initial preparation of the internal state of the ions. This is a very useful feature when we think about the use of such a system for the implementation of quantum information tasks. For example, consider that all N ions are initially prepared in their ground state $|g\rangle^{\otimes N}$. In this case, the effective Hamiltonian in interaction picture will be reduced to
\begin{equation}
\hat{H}_{{\rm{eff}}}^{|g\rangle^{\otimes N}}=-\sum_{i=1}^N\frac{g_i^2}{\Delta{i}}f^2(\hat{a}_i^\dag\hat{a}_i)\hat{b}^\dag\hat{b}.
\label{g}
\end{equation}
This Hamiltonian describes a cross-Kerr interaction between the vibrational bosonic operators of any element of the cluster of ions with the bosonic field operators. The single ion case has been previously studied in \cite{kerr} where a quantum phase gate robust against spontaneous emission has been proposed. Again, it is noteworthy that the physical mechanism responsible for the appearance of the cross-Kerr interaction between motion and field is the dynamical atomic Stark-shifts induced by the quantized cavity field. As already said before, we will analyze the entanglement generated due to this coupling in Section \ref{sta}.
\section{NLMS generation protocol}\label{gen}
We now propose a scheme for generating entangled even and odd states of the form
\begin{equation}\label{psi}
|\Psi_+\rangle=\frac{1}{N_{\psi_+}}(|C_+,C_+\rangle+|C_-,C_-\rangle)
\end{equation}
or 
\begin{equation}\label{phi}
|\Phi_+\rangle=\frac{1}{N_{\phi_+}}(|C_+,C_-\rangle+|C_-,C_+\rangle),
\end{equation}
where
$|C_\pm\rangle=(|\alpha\rangle\pm|-\alpha\rangle)/N_{\pm}$, with $|\pm\alpha\rangle$ being single-mode coherent states and $N_{\pm,\psi_+,\phi_+}$ the normalization constants. The entangled states (\ref{psi}) and (\ref{phi}) are also commonly called nonlocal mesoscopic states NLMS in the case they involve two well spatially separate modes, and they are very suitable for the study of nonlocal properties of quantum mechanics. Actually, the states $|C_+\rangle$ and $|C_-\rangle$ are eigenstates of the parity operator $e^{i\pi\hat{a}^\dag\hat{a}}$, and this fact implies that the states (\ref{psi}) and (\ref{phi}) involve the entanglement of different parities for the bosonic modes. This is the key ingredient for testing non-locality in these two-mode entangled states as shown in \cite{desbell}. They propose the use of a adapted version of a simple Bell inequality that may be violated by the NLMS for carefully chosen points in phase space. Although, there are some other proposals in the literature for generating NLMS in the context of cavity quantum electrodynamics \cite{haroche,genprop1} or trapped ions \cite{genprop2}, we consider here the system comprising trapped ions and quantized fields altogether, and this brings out a quite different way for accomplishing the generation of NLMS. This new scheme makes use of the atomic dipole-dipole coupling that depends on the intensity of the motion of the trapped ions (\ref{dip}). Due to the combined effect of the Rabi oscillations with the motional dependent coupling, the whole system naturally evolves to a multipartite entangled state that, after postselection, collapses to the desired NLMS involving ions in different traps, as we are going to show next. Let us consider now the case $N=2$ in (\ref{eff}) and also that the electronic initial state of one ion has been prepared in excited state $|e_1\rangle$ and the other in the ground state $|g_2\rangle$. Just like before, the effective Hamiltonian (\ref{eff}) reduces to a specific form depending on the internal electronic state preparation, and for the case now considered it reduces to
\begin{eqnarray}\label{eff2}
\hat{H}_{{\rm{eff}}}^{|e_1\rangle|g_2\rangle}&=&\nu_1\hat{a}_1^\dag\hat{a}_1+\nu_2\hat{a}_2^\dag\hat{a}_2+\omega_c\hat{b}^\dag\hat{b}+\omega_1\frac{\hat{\sigma}_1^z}{2}+\omega_2\frac{\hat{\sigma}_2^z}{2}+(1+\hat{b}\hat{b}^\dag)\left[\frac{g_1^2}{\Delta{1}}f^2(\hat{a}_1^\dag\hat{a}_1)\hat{\sigma}_1^+\hat{\sigma}_1^-+ \frac{g_2^2}{\Delta{2}}f^2(\hat{a}_2^\dag\hat{a}_2)\hat{\sigma}_2^+\hat{\sigma}_2^-\right]\nonumber\\ &&+\frac{g_1g_2}{2}\left(\frac{\Delta_1+\Delta_2}{\Delta_1\Delta_2}\right)f(\hat{a}_1^\dag\hat{a}_1)f(\hat{a}_2^\dag\hat{a}_2)(\hat{\sigma}_1^+\hat{\sigma}_2^-+\hat{\sigma}_1^-\hat{\sigma}_2^+).
\end{eqnarray} 
If we consider now the two ions to be identical, i.e, $\Delta_1=\Delta_2=\Delta$, the Hamiltonian (\ref{eff2}) will be written in the interaction picture as
\begin{eqnarray}\label{eff3}
\hat{H}_{{\rm{eff}}}^{|e_1\rangle|g_2\rangle}&=& \frac{g_1g_2}{\Delta}f(\hat{a}_1^\dag\hat{a}_1)f(\hat{a}_2^\dag\hat{a}_2)(\hat{\sigma}_1^+\hat{\sigma}_2^-+\hat{\sigma}_1^-\hat{\sigma}_2^+).
\end{eqnarray} 
We note that the Hamiltonian (\ref{eff3}) is independent of the state of the cavity field. It follows that applications relying upon (\ref{eff3}) will be robust against $T=0$ cavity losses and would work well even for ideal cavities sustaining a thermal field \cite{shibiao}.

The NLMS generation protocol starts with the preparation of the system's initial state
\begin{equation}\label{in}
|\psi(0)\rangle=|e_1,\alpha_1,g_2,\alpha_2\rangle,
\end{equation}
i.e, the vibrational degree of freedom of the ion $1(2)$ is prepared in the coherent state $|\alpha_{1(2)}\rangle$. We would like to point out that such initial preparation is quite feasible in the actual experimental status of laser manipulated trapped ions \cite{review}. As a matter of fact, the experimental realization of vibrational coherent states for a $^9$Be$^+$ ion has already been reported \cite{cohgen}. 

According to (\ref{eff3}) the initial state (\ref{in}) evolves to 
\begin{equation}\label{ev1}
|\psi(t)\rangle=|e_1,g_2\rangle\cos(\hat{\Omega}t)|\alpha_1,\alpha_2\rangle-i|g_1,e_2\rangle\sin(\hat{\Omega}t)|\alpha_1,\alpha_2\rangle,
\end{equation} 
where
\begin{equation}\label{rop}
\hat{\Omega}=\frac{g_1g_2}{\Delta}f(\hat{a}_1^\dag\hat{a}_1)f(\hat{a}_2^\dag\hat{a}_2).
\end{equation}
We now assume the Lamb-Dicke regime ($\eta_{1,2}\ll 1$) where the motion of the ions is restrict to a spatial extent much smaller than the cavity field wavelength. For a further simplification, let us also take $\eta_1=\eta_2=\eta$, and then the operator (\ref{rop}) may be approximated by
\begin{equation}
\hat{\Omega}\approx\frac{g_1g_2}{\Delta}\left(1-\eta^2-\eta^2[\hat{a}_1^\dag\hat{a}_1-\hat{a}_2^\dag\hat{a}_2]\right).
\end{equation}
In this approximation, the evolved state of the system (\ref{ev1}) will assume the form
\begin{eqnarray}\label{state1}
|\psi(t)\rangle &=&|e_1,g_2\rangle\{\cos(\omega_\eta t)[|\phi_1^+,\phi_2^+\rangle+|\phi_1^-,\phi_2^-\rangle]-i\sin(\omega_\eta t)[|\phi_1^-,\phi_2^+\rangle+|\phi_1^+,\phi_2^-\rangle]\}\nonumber\\
&&+|g_1,e_2\rangle\{\cos(\omega_\eta t)[|\phi_1^-,\phi_2^+\rangle+|\phi_1^+,\phi_2^-\rangle]-i\sin(\omega_\eta t)[|\phi_1^+,\phi_2^+\rangle+|\phi_1^-,\phi_2^-\rangle]\},
\end{eqnarray}
where $\omega_\eta=g_1g_2(1-\eta^2)/\Delta$, and
\begin{eqnarray}
|\phi_j^{+}\rangle &=&\frac{|\alpha_je^{i\frac{\eta^2g_1g_2}{\Delta}t}\rangle+|\alpha_je^{-i\frac{\eta^2g_1g_2}{\Delta}t}\rangle}{2}\nonumber\\
|\phi_j^{-}\rangle &=&\frac{|\alpha_je^{i\frac{\eta^2g_1g_2}{\Delta}t}\rangle-|\alpha_je^{-i\frac{\eta^2g_1g_2}{\Delta}t}\rangle}{2i},
\end{eqnarray}
with $j=1,2$. Now, taking $\alpha_1=\alpha_2=\alpha$, and an interaction time $\frac{\eta^2g_1g_2}{\Delta}t_I=\pi/2$, the state (\ref{state1}) takes the form
\begin{eqnarray}
|\psi(t_I)\rangle\propto|\Psi_+\rangle\{\cos\theta_\eta|e_1,g_2\rangle
-i\sin\theta_\eta|g_1,e_2\rangle\}+|\Phi_+\rangle\{\cos\theta_\eta|g_1,e_2\rangle
-i\sin\theta_\eta|e_1,g_2\rangle\},
\label{state2}
\end{eqnarray}
with $\theta_\eta=\frac{(1-\eta^2)\pi}{2\eta^2}$. Now we will explain the post-selection procedure that will allow us to obtain the NLMS in the form (\ref{psi}) and (\ref{phi}) for the motion of the trapped ions. 
 If the Lamb-Dicke parameter is carefully set to fulfill the condition $\theta_\eta=k\pi/2$, i.e. $\eta=\sqrt{1/(1+2k)}$, with the $k$ integer or half-integer, the state (\ref{state2}) reduces to either
\begin{eqnarray}
|\psi(t_I)\rangle &\propto &|\Psi_+\rangle|e_1,g_2\rangle+|\Phi_+\rangle|g_1,e_2\rangle,
\end{eqnarray} 
for the $k$ integer, and
\begin{eqnarray}
|\psi(t_I)\rangle &\propto &|\Psi_+\rangle|g_1,e_2\rangle+|\Phi_+\rangle|e_1,g_2\rangle,
\end{eqnarray} 
for the $k$ half-integer. Then, after choosing $k$, a simple measurement on the internal state of the ions would lead to the generation of the states (\ref{psi}) and (\ref{phi}). The experimental
discrimination between the two internal electronic levels may be done using
the very efficient electron shelving method \cite{shelving,review}. Following our proposal, one could then generate NLMS involving the state of well separate trapped ions and then perform the experimental investigation of its nonlocal properties using the proposals described in \cite{desbell}. Finally, we would like to mention that there are other different mechanisms and regimes used to generate entanglement in this system \cite{peter,otherions}.

\section{Entanglement study}\label{sta}
We now present a study of the entanglement properties of our model. When the cluster of ions is initially prepared in the ground electronic state $|g\rangle^{\otimes N}$, the resulting Hamiltonian is the cross-Kerr interaction (\ref{g}). It involves the coupling between the vibrational motion of any ion in the cluster and the cavity field. Just like before, we will be interested in the common Lamb-Dicke regime characterized by the condition $\eta_i\ll 1$. In this regime, the Hamiltonian (\ref{g}) will read
\begin{equation}\label{g2}
\hat{H}_{{\rm{eff}}}^{|g\rangle^{\otimes N}}=-\sum_{i=1}^N\lambda_i\hat{a}_i^\dag\hat{a}_i\hat{b}^\dag\hat{b},
\end{equation}
with $\lambda_i=2\eta^2g_i^2/\Delta_i$. We remark that the natural frequencies of all oscillators have been shifted and included in the free Hamiltonian not shown \cite{kerr}. 

In order to study the dynamics of entanglement in our model (\ref{g2}), we employ the linear entropy \cite{linear} as a bipartite entanglement measure to analyze different partitions of the system, and the global entanglement(GE) \cite{global} to discuss the multipartite entanglement between all elements of the cluster and the cavity field. The linear entropy and the global entanglement have been used quite extensively in the investigation of entanglement behavior in quantum phase transitions \cite{phaset,applic}. 
 
In what follows we investigate the entanglement dynamics for three types of initial states of physical interest. First we consider both the ions and the field in {\sl coherent} states (case A),  then the ions in {\sl coherent} states and the field in a {\sl squeezed vacuum} state (case B), and finally the ions in {\sl squeezed vacuum} states and the field in a {\sl coherent} state (case C).

A) {\sl Initially coherent states}: let us consider the following initial state $|\Psi(0)\rangle = \otimes_{j=1}^n|\alpha_j \rangle \otimes|\alpha_0 \rangle$.
Since the interaction is diagonal, the time-evolved state $|\Psi(t)\rangle = U(t)|\Psi(0)\rangle$ can be easily calculated and the reduced density operators for one of the ions ($k$-th, $k=1,..N$) and the field (tracing over all the other degrees of freedom) can be exactly calculated. The linear entropy for the $a$-th subsystem can be calculated by means of the expression $S_L(\rho_a)= [1- Tr_{a} \rho_a^2]$ ($\rho_a$ is the reduced density operator of the $a$-th subsystem $a=0,..N$, where all the other degrees of freedom have been traced out) giving for the field ($a=0$) and the $k$-th ion entanglement the expressions below
\begin{eqnarray}
S_L(\rho_0(t))=1-\sum_{n,m}P_n^c(\alpha_0)P_m^c(\alpha_0) e^{-\sum_{j=1}^N 2|\alpha_j|^2(1-\cos [\lambda_j(m-n)t])}
\label{flentropy}
\end{eqnarray}
\begin{eqnarray}
S_L(\rho_k(t))=1-\sum_{n,m}P_n^c(\alpha_0)P_m^c(\alpha_0) e^{- 2|\alpha_k|^2(1-\cos [\lambda_k(m-n)t])}, 
\label{ilentropy}
\end{eqnarray}
where $P_n^c(\alpha_0)=e^{-|\alpha_0|^2}\frac{|\alpha_0|^{2n}}{n!}$ and $k=1,..N$.

\begin{figure}[h]
\begin{center}
\vspace{-0.5cm}
\hspace{-2cm}
\includegraphics[scale=0.40,angle=-90]{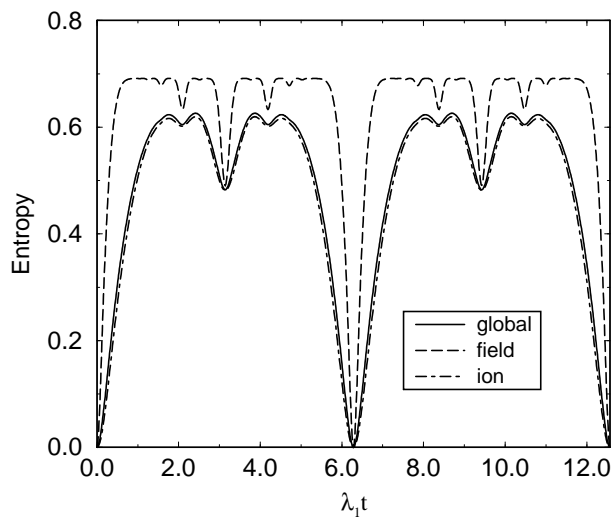}
\end{center}
\caption{Plots of entanglements as a function of time for all coherent initial state: global (continuous), field (dotted lines), and individual ion (dot-dashed lines) ; for the following parameter choices $N=10$, $\lambda_k=1.0\lambda_1$ ($k=2,..,10$) , $\alpha_i=\alpha_0=1.0$.}
\label{fig1}
\end{figure}

\begin{figure}[h]
\begin{center}
\vspace{-0.5cm}
\includegraphics[scale=0.40,angle=-90]{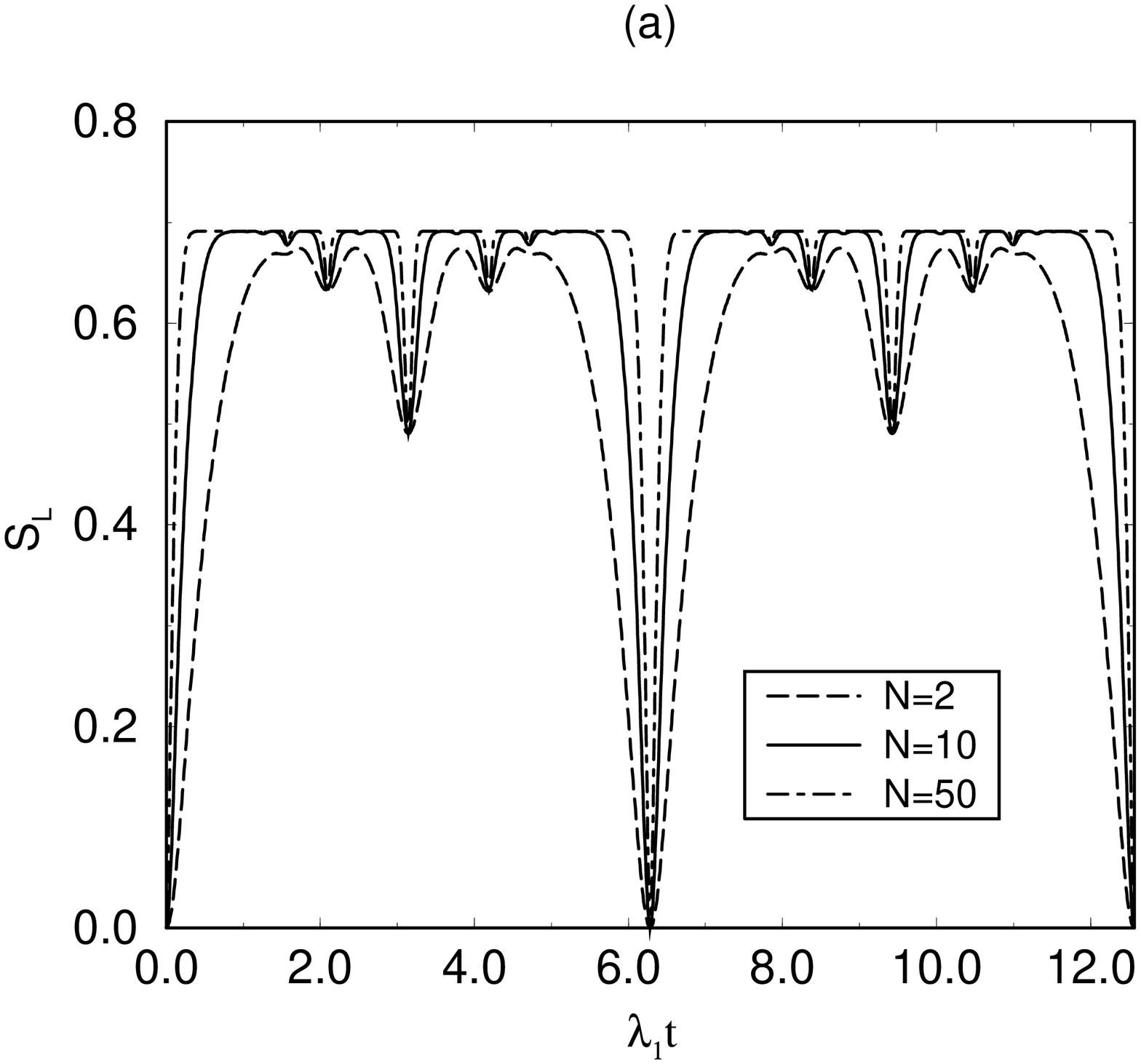}
\end{center}
\includegraphics[scale=0.40,angle=-90]{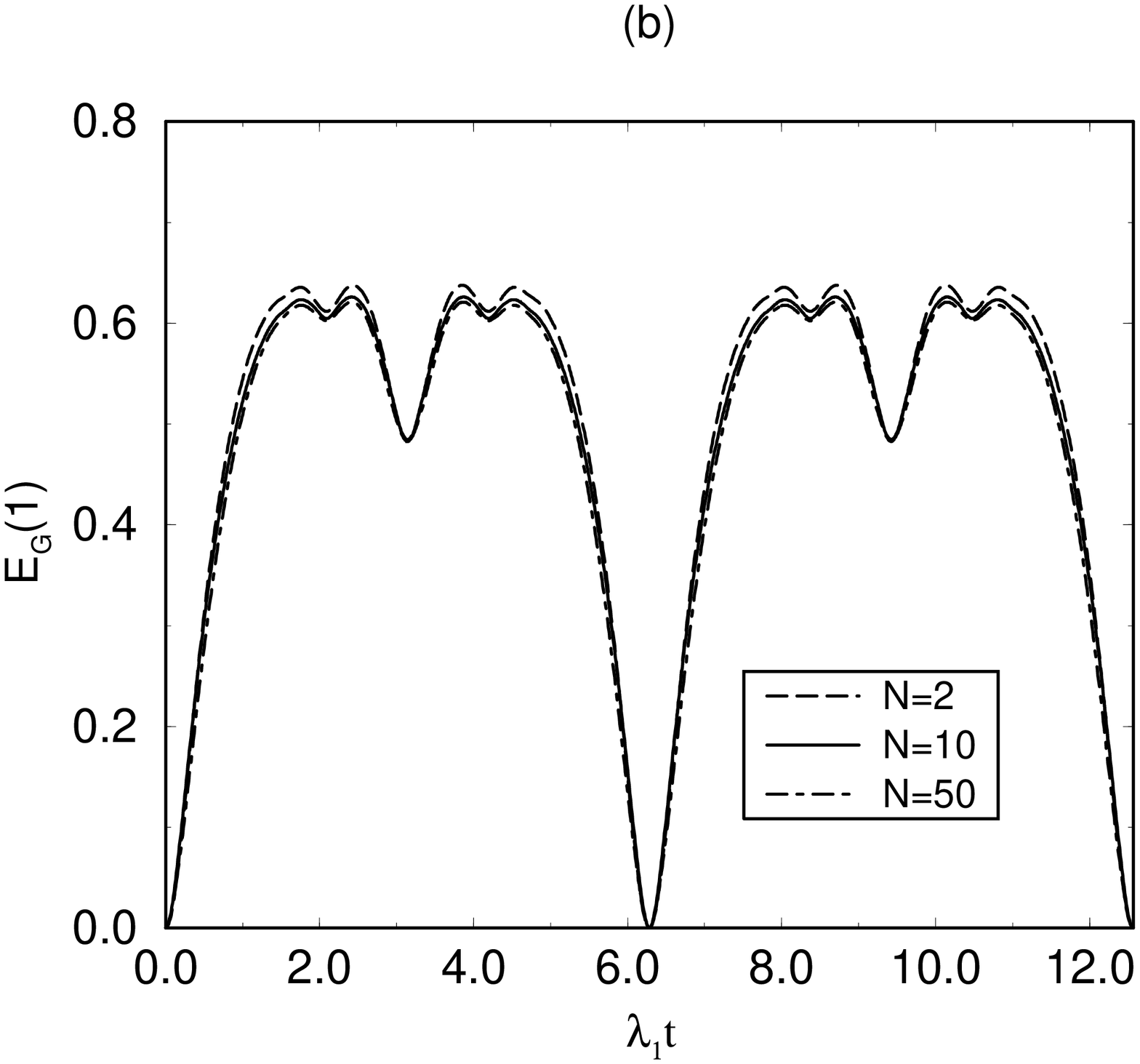}
\caption{Entanglements as a function of time for all coherent initial state. (a) field, (b) global; for the following parameter choices $N=2,10,50$, $\lambda_k=1.0\lambda_1$ ($2=1,..N$) , $\alpha_i=\alpha_0=1.0$.}
\label{fig2}
\end{figure}

\begin{figure}
\begin{center}
\vspace{-1cm}
\hspace{-1cm}
\includegraphics[scale=0.4,angle=-90]{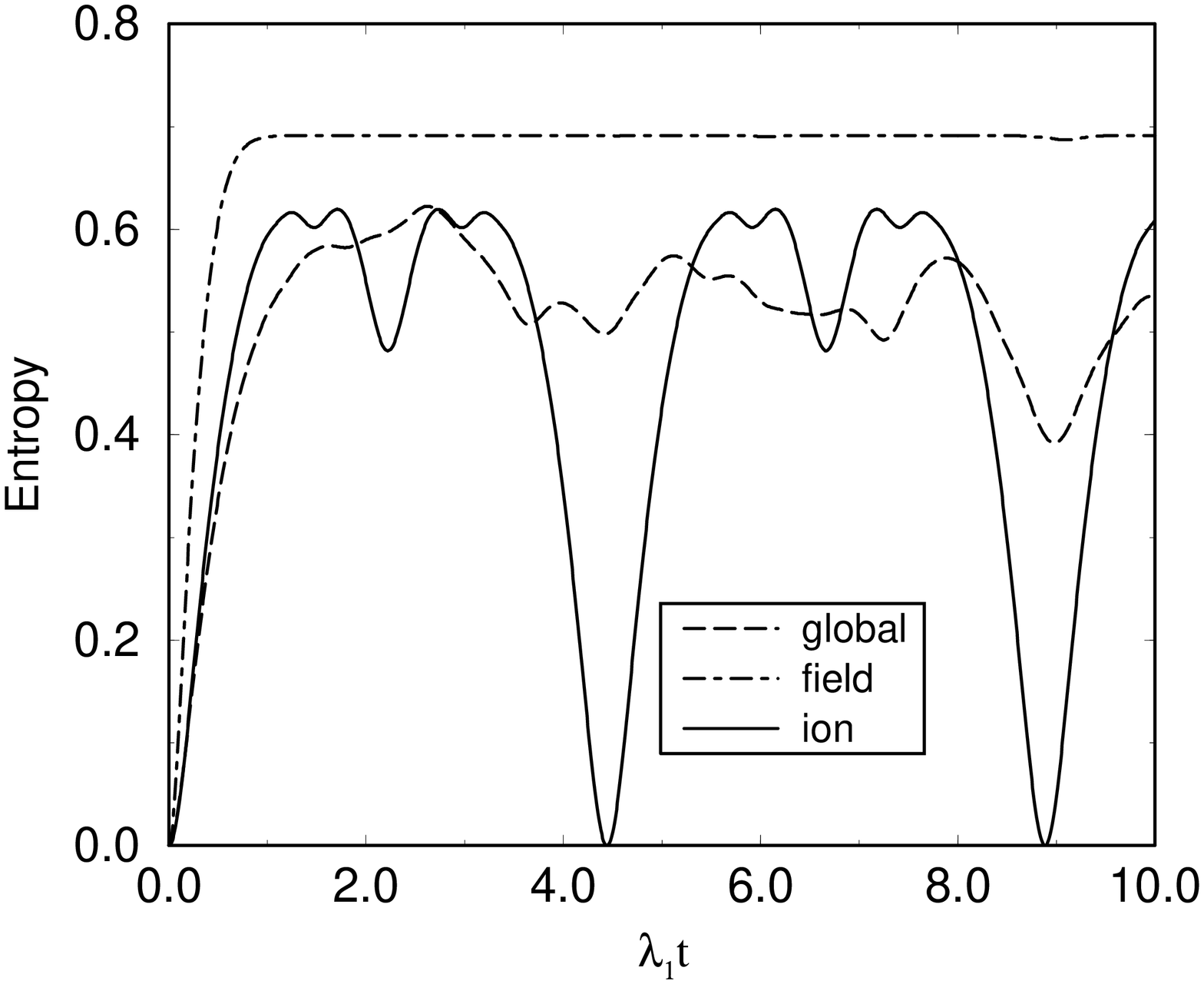}
\end{center}
\caption{Entanglements as a function of time for the all coherent initial state: global (dashed), field (dot-dashed lines) and ion-2 (continuous line); for the following parameter choices $N=5$, $\lambda_2= \sqrt{2}\lambda_1,\lambda_3=1/\sqrt{2}\lambda_1,\lambda_4=\sqrt{3}\lambda_1,\lambda_5=1/\sqrt{3}\lambda_1$, $\alpha_i=\alpha_0=1.0$.}
\label{fig3}
\end{figure}

We also consider the global entanglement measure in the form defined by Oliveira et al. \cite{applic} which for a system with $M$ identical parties is given by
\begin{eqnarray}
E_G(1)= \left(\frac{d}{d-1}\right)\frac{1}{M}\sum_{a}S_L(\rho_a),
\label{ge}
\end{eqnarray}
where $S_L(\rho_a)$ is the linear entropy of the $a$-th individual subsystem and $d$ is the dimensionality of the each subspace. In our case, $M=N+1$ and all subsystems are harmonic oscillators $(d=\infty)$ leading to
\begin{widetext}
\begin{eqnarray}
E_G(1)&=&\frac{1}{N+1}\sum_{a=0}^N S_L(\rho_a)\nonumber \\
 &=& 1-\sum_{n,m} \frac{P_n^c(\alpha_0) P_m^c(\alpha_0)}{N+1} \left(\sum_{j=1}^{N} e^{ - 2|\alpha_j|^2(1-\cos [\lambda_j(m-n)t])}- e^{-\sum_{j=1}^N 2|\alpha_j|^2(1-\cos [\lambda_j(m-n)t])}\right).
\label{geA}
\end{eqnarray}
\end{widetext}
In Fig.\ref{fig1}, we plot the case with equal interaction strength of the ions with the field, $\alpha_k=\alpha_0=1$, and $N=10$. In this equal coupling case, all the entanglements are very similar, with the field one slightly higher than the individual ion's (from Eq.(\ref{ilentropy}) all ions have identical behavior). The global entanglement behaves like the individual entanglements.
In Fig.\ref{fig2}(a), we show that the field remains at a plateau of entanglement for longer and longer times as we increase the number of ions ($N=2,10,50$), except at those times at which it disentangles or partially disentangles. On the other hand, the global entanglement changes little with $N$ as shown in Fig.\ref{fig2}(b).  
In Fig.\ref{fig3}, we plot the case of unequal couplings with $N=5$, where the incommensurability of the coupling constants causes the global entanglement curve to become irregular and non-periodic. It is interesting to notice that in this case the field remains in the plateau and does not return to zero. This happens because the field is always entangled with some of the ions, although individual ions present periodic behavior (entanglement of ion-$2$ is shown in this case).  
 
\begin{figure}
\begin{center}
\includegraphics[scale=0.4,angle=-90]{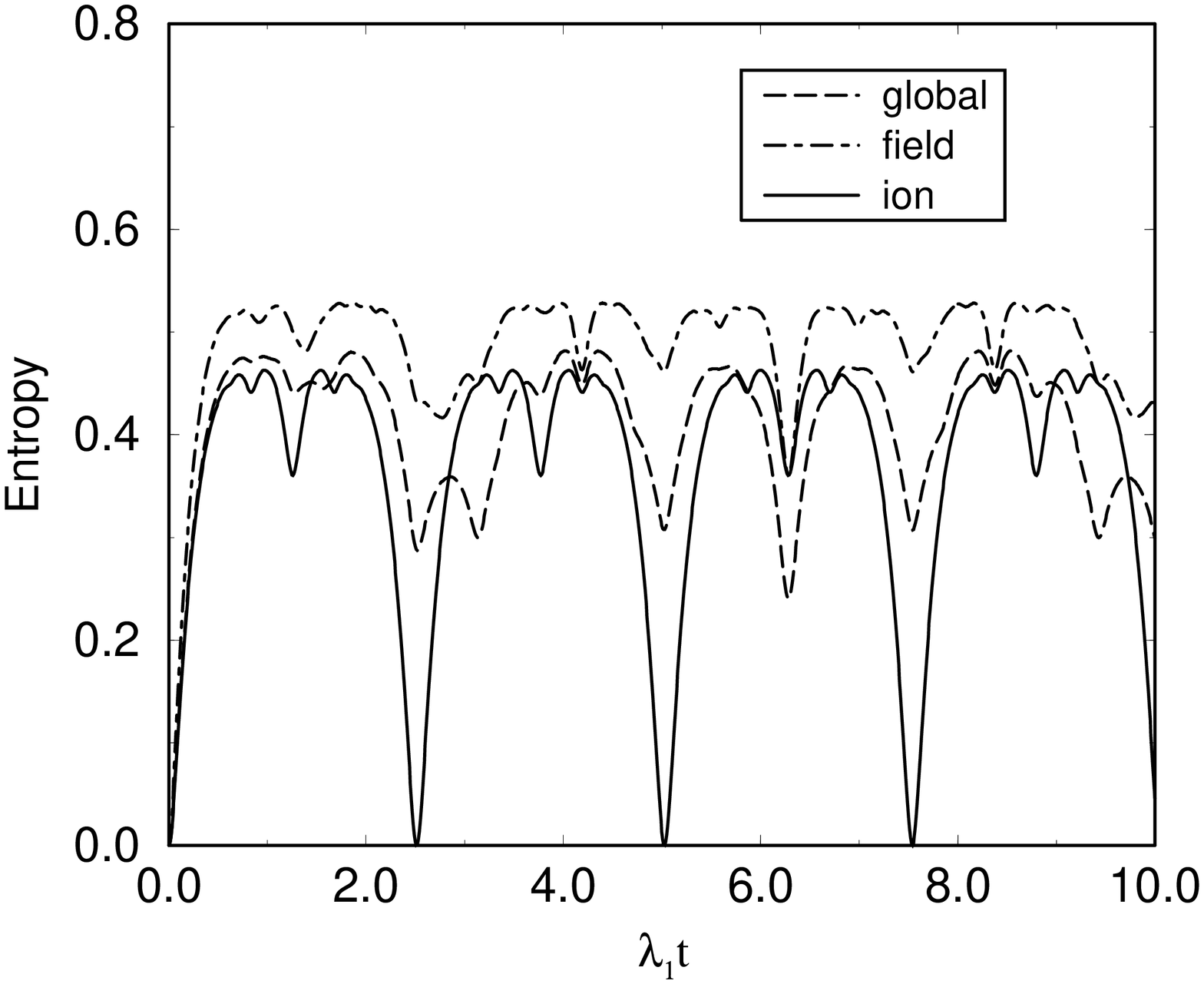}
\end{center}
\caption{Plots of entanglements as a function of time for initially {\sl squeezed field} and coherent ions: global (dashed), field (dot-dashed lines), ion-2 (continuous line),; for the following parameter choices $N=2$, $\lambda_2=1.25\lambda_1$ , $\alpha_i=r_c=1.0$.}
\label{fig4}
\end{figure}

  B) Initially {\sl squeezed vacuum state for the field}: let us consider the initial state $|\Psi(0)\rangle=\otimes_{j=1}^n|\alpha_j \rangle \otimes |0,\xi_f \rangle$, with $\xi_f=r_f e^{i \varphi_f}$. For this initial state we get a similar expression for the linear entropies of the field (\ref{flentropy}) and $k$-th ion (\ref{ilentropy}), with the summation done only for even values of $n,m$ and the distribution function $P_n^c$ substituted by the one for squeezed state 
\begin{center} 
\begin{equation*}
P_n^s(r_f)= \frac{1}{\cosh (r_f)}\frac{n!}{\left[(\frac{n}{2})!\right]^2}\left[\frac{\tanh(r_f)}{2}\right]^n
\end{equation*} 
\end{center} 
($n$ even), where we have chosen $\varphi_f=0$. Therefore the GE in this case is also given by (\ref{geA}) with the same substitutions. In Fig.\ref{fig4}, we plot the ion-2, field and global entanglements for $N=2$ and unequal couplings with the field. The squeezing of the initial field state lowers the maximum values of the entanglement as compared with the coherent field state (see Fig.\ref{fig2}(a)). Also, due to the fact that the couplings with the field are again different, the field and global entanglement exhibits non-periodic behavior as noticed in the previous cases (see Fig.\ref{fig3}). This allows one to engineer a decoherence scenario via cross-Kerr type interaction \cite{renato} of the field by choosing appropriate values of the parameter $\lambda_i= 2 \eta^2 g_i /\Delta_i$ for each ion. The easiest way of doing this modulation is by choosing the detunings appropriately through the application of external electric fields on each ion (static Stark effect). 

C) Initially {\sl squeezed vacuum states for the ions}: consider now all the ions initially squeezed $|\Psi(0)\rangle=\otimes_{j=1}^n |0,\xi_j \rangle \otimes|\alpha_f \rangle$ with $\xi_j=r_j e^{i \varphi_j}$ (for simplicity $\varphi_j=0$ for all $j$). The field and ion linear entropies result as
\begin{eqnarray}
S_L(\rho_f(t))=1-\Pi_{j=1}^N\sum_{{\rm{even}}\,\, n,m}P_n^s(r_j)P_m^s(r_j) e^{2|\alpha_0|^2(\cos [ \lambda_j(m-n)t]-1)},
\label{flentropy3}
\end{eqnarray}
\begin{eqnarray}
S_L(\rho_k(t))=1-\sum_{{\rm{even}}\,\, n,m}P_n^s(r_k)P_m^s(r_k) e^{2|\alpha_0|^2(\cos [ \lambda_k (m-n)t]-1)}.
\label{ientropy3}
\end{eqnarray}
The corresponding GE is given by the following expression
\begin{eqnarray}
E_G(1) = 1-\frac{1}{N+1}\sum_{{\rm{even}}\,\, n,m}[ \Pi_{j=1}^N P_n^s(r_j) P_m^s(r_j)+\sum_{j=1}^N  P_n^s(r_j) P_m^s(r_j)]e^{2|\alpha_0|^2(\cos [ \lambda_j(m-n)t]-1)}
\label{geC}
\end{eqnarray}
%
\begin{figure}
\begin{center}
\includegraphics[scale=0.4,angle=-90]{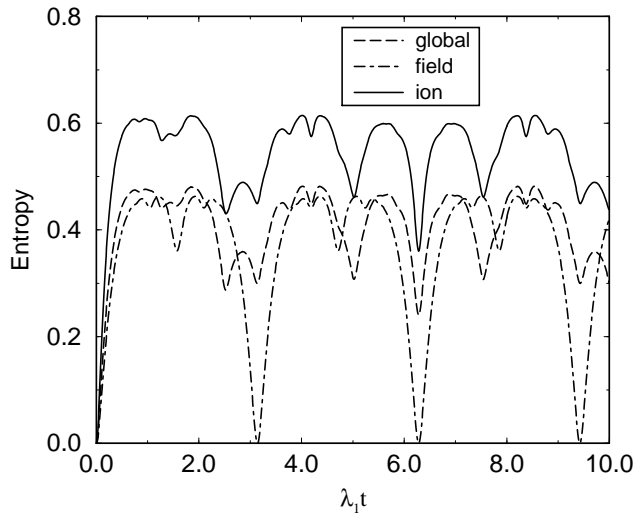}
\end{center}
\caption{Plots of entanglements as a function of time for initially {\sl squeezed ions} and coherent field: global (dashed), field (dot-dashed lines), and ion-2 (continuous line), for the following parameter choices $N=2$, $\lambda_2=1.25\lambda_1$ , $\alpha_0=r_1=r_2=1.0$.}
\label{fig5}
\end{figure}

In Fig.(\ref{fig5}) we show the time evolution of the linear entropy of the ion-2, field and the global entanglement. It is notable that the field and individual ion entanglements exchange the role from Fig.\ref{fig4} to Fig.\ref{fig5}, namely that the field entanglement is now {\sl smaller} than the entanglements of the individual ions. Also, it becomes clear that in the case where either the field or the ions are in the squeezed states, the subsystem which is coherent shows a typical periodic behavior, whereas the ones which are in the squeezed states have a more irregular behavior. 
\section{Conclusion} \label{concl}

We have investigated the dynamics of a cluster of N trapped ions in the dispersive interaction with a quantized electromagnetic field. Depending on an appropriate preparation of the internal state of the ions in the cluster, we were able to obtain either cross-Kerr type interactions between field and ions or ionic dipole-dipole couplings that depend on the intensity of the motion of the trapped ions. With such dipole-dipole coupling, simple postselection, and appropriate choice of the interaction time, one is able to generate non-local entangled mesoscopic states (NLMS) of two well separated ions, as we have shown.

A simple preparation of all ions in the ground state has led to motion-dependent Stark-shifts in the form of cross-Kerr type interactions between field and ions. This allowed us to study both the bipartite entanglement of the field with the cluster or each of the ions with the rest of the system, and {also to evaluate the multipartite global entanglement. We have considered coherent and squeezed initial states which are of interest in many physical situations and seems to be practically feasible in the current status of laser manipulated trapped ions experiments. In the case of a large cluster of ions, we analyze the possibility of observing controlled decoherence of the field in laboratory by choosing both an appropriate initial set of states for the system and an adequate set of effective interactions via detunings to dephase the periodicity of the entanglement of individual ions.

\begin{acknowledgments}
This work is partially supported by CNPq (Conselho Nacional de Desenvolvimento Cient\'\i fico e Tecnol\'ogico) (KF), and FAPESP (Funda\c c\~ao de Amparo \`a Pesquisa do Estado de S\~ao Paulo) (FLS) grant number 05/04533-7, Brazil.
\end{acknowledgments}


\begin{thebibliography}{99}
\bibitem{foot1}A multiplicative factor has been adjusted to give the correct coupling constant for the effective Hamiltonian consistent with the other Heisenberg equations, see \cite{puri}.
\end{thebibliography}
\end{document}